\documentclass[prl, twocolumn,groupedaddress]{revtex4}
\usepackage{graphicx}
\usepackage{amsmath}
\usepackage{amssymb}
\usepackage{bm}
\usepackage{color}
\usepackage{hyperref}
\usepackage{tabularx}
\usepackage{multirow}
\usepackage{braket}

\begin{document}
\title{Zeeman-Induced Gapless Superconductivity with Partial Fermi Surface}
	
\author{{ Noah F. Q. Yuan $ ^{1,2} $ and Liang Fu $ ^1 $}}\thanks{Correspondence address : liangfu@mit.edu}
\affiliation{1. Department of Physics, Massachusetts Institute of Technology, Cambridge,
Massachusetts 02139, USA\\
2. Department of Physics, Hong Kong University of Science and Technology, Clear Water Bay, Hong Kong, China  }

\begin{abstract}
We show that an in-plane magnetic field can drive two-dimensional spin-orbit-coupled systems under superconducting proximity effect into a gapless phase where parts of the normal state Fermi surface are gapped, and the ungapped parts are reconstructed into a small Fermi surface of Bogoliubov quasiparticles at zero energy. Charge distribution, spin texture, and density of states of such ``partial Fermi surface'' are discussed. Material platforms for its physical realization are proposed.  
\end{abstract}

\maketitle
{\bf Introduction}---
In recent years spin-orbit coupling (SOC) is found to play an increasingly important role in experimental and theoretical studies of superconductivity. The Rashba SOC is a key ingredient in creating Majorana bound states via superconducting proximity effect \cite{FuKane,Alicea,Oreg,Potter,Kouwenhoven,ADas,Lutchyn,JFJia}. The Ising-type SOC can stablize two-dimensional (2D) superconductivity against very large in-plane magnetic fields \cite{JMLu,Saito,Xi,BTZ}. Strong atomic SOC can enhance $p$-wave pairing in inversion-symmetric metals, leading to time-reversal-invariant topological superconductivity \cite{FuBerg,Sho,KoziiFu,YXWang, WuMartin}.
The interplay between SOC and superconductivity continues to be a fruitful source of new physics.

Rashba SOC also brings new twists to 2D superconductors under an in-plane magnetic field that couples to electron spin. The Zeeman energy is pair breaking for $s$-wave superconductivity. In the absence of SOC, a transition from superconducting to normal state should occur when the Zeeman splitting exceeds the superconducting condensate energy \cite{Clogston,Chandrasekhar,Maki}. For 2D superconductors with strong Rashba SOC, recent works \cite{Gorkov,Olga,Patrick} proposed that superconductivity with finite-momentum pairing may be stabilized at high field, thus leading to the Fulde-Ferrell-Larkin-Ovchinnikov state.

In this work, we study the effect of Zeeman field on 2D spin-orbit-coupled electron systems, which are by themselves non-superconducting but acquire a superconducting gap by proximity coupling to an external superconductor. Such systems include---but are not limited to---superconductor-topological insulator (TI) \cite{MXWang,SYXu,Hart,Yazdani} and superconductor-InAs 2DEG hybrid structures \cite{Marcus0,Marcus1,Marcus2}, where a hard proximity-induced superconducting gap at zero field has been observed. We show that an increasing in-plane field can close the proximity-induced gap and create gapless Bogliubov quasiparticles before eventually destroying the parent superconductor. This scenario is realized when the $g$-factor of the 2D system is sufficiently larger than that of the parent superconductor, or when the proximity-induced gap is sufficiently smaller than the parent superconducting gap. Interestingly, in the presence of strong Rashba SOC, the competition between proximity-induced pairing at zero total momentum and pair-breaking Zeeman field partially gaps the electron Fermi surface, and reconstructs the ungapped segments 
into a banana-shaped Fermi surface of zero-energy Bogoliubov quasiparticles. These Bogoliubov quasiparticles are coherent superpositions of electrons and holes residing on two arcs on opposite sides of the original Fermi surface. We call the  Bogoliubov Fermi surface ``partial Fermi surface''. We discuss charge distribution, spin texture and density of states of such partial Fermi surface and generalize these results to 2D superconductors with spin-nondegenerate Fermi surfaces exhibiting arbitrary spin texture under in-plane Zeeman field.

\begin{figure}
\centering
\includegraphics[width=3.20in]{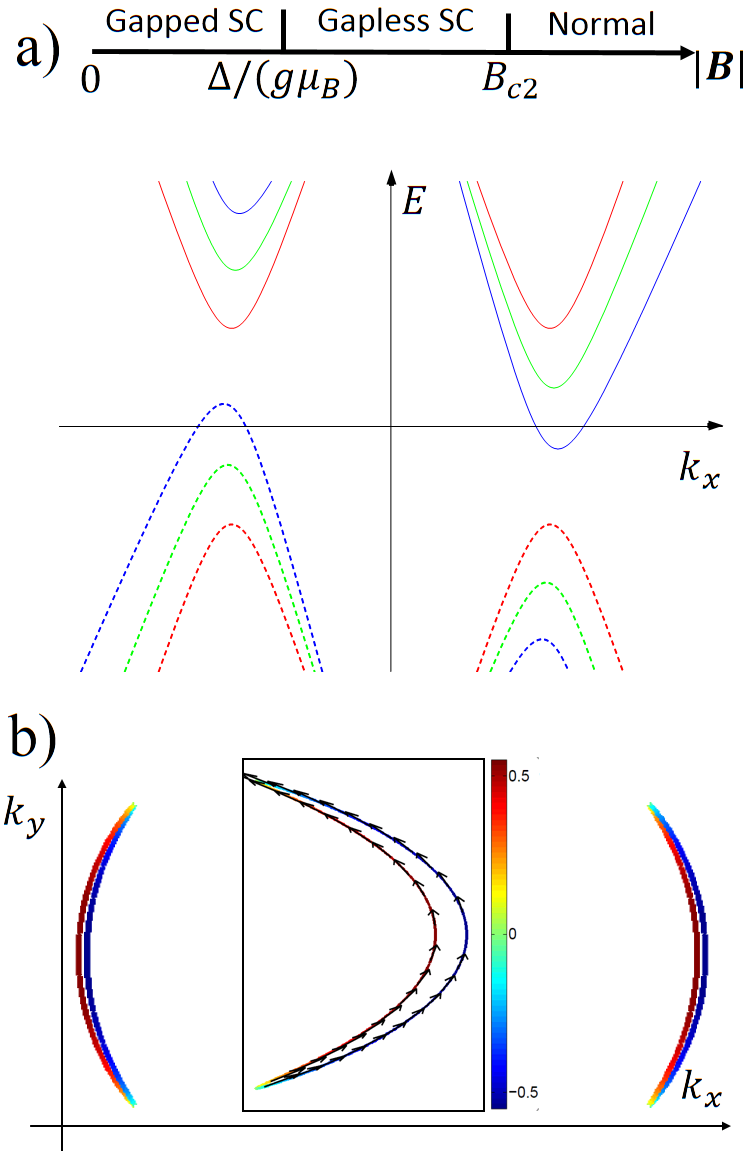}
\caption{a) Upper: The phase diagram of superconducting topological insulator (TI) surface states described by Hamiltonian (\ref{H}), in terms of magnetic field $|\bm B|$. When $ |\bm B|>B_{c2} $ the whole system is in normal phase. The superconducting (SC) phase in the region $ \frac{\Delta}{g\mu_B}<|\bm B|<B_{c2} $ is gapless while the SC phase in $ 0<|\bm B|<\frac{\Delta}{g\mu_B} $ is fully gapped, with $g$-factor $ g $ and Bhor magneton $\mu_B$. Lower: Energy spectra of Hamiltonian (\ref{H}) when $ k_y=0 $. The red, green and blue curves correspond to $ V=0,0.6,1.2 $ respectively, the solid and dashed lines correspond to quasiparticles and quasiholes respectively. For illustration the parameters are $ v_F =10, \mu =5,\Delta =1 $. b) The partial Fermi surfaces in the gapless SC phase where $ v_F =100, \mu =50,V=1.2,\Delta =1 $. Colors denote the charge distribution in unit of $e$. Inset: The zoom-in plot of partial Fermi surface on the $k_x>0$ side where the arrows indicate the spin polarization.}\label{1}
\end{figure}

{\bf Surface States of Topological Insulators}---
We first consider the TI surface states in proximity with an $s$-wave superconductor \cite{FuKane} and under an in-plane magnetic field $\bm B$. The Hamiltonian reads
\begin{eqnarray}\label{H}
H&=&\sum_{\bm k}c^{\dagger}_{\bm k}[v_{F}(k_{x} \sigma_{y}- k_{y}\sigma_{x}) -\mu -V  \sigma_{y}]c_{\bm k} \nonumber \\
&+& \Delta( c^{\dagger}_{\bm k\uparrow}c^{\dagger}_{-\bm k\downarrow} +h.c.),
\end{eqnarray}
where $ c^{\dagger}_{\bm k}=(c^{\dagger}_{\bm k\uparrow},c^{\dagger}_{\bm k\downarrow}) $ is electron creation operator at momentum $\bm k$ and with spin $s=\uparrow,\downarrow$. Here $v_{F}$ is Fermi velocity of surface states. $\mu$ is the chemical potential measured from the Dirac point. $\Delta$ is the induced $s$-wave pairing potential, which is generally smaller than the gap of the parent superconductor. $ V =g\mu_B |\bm B| $ is the Zeeman energy induced by in-plane field $\bm B$ with $g$-factor $g$ and Bohr magneton $\mu_B$. Without loss of generality, we choose $\bm B$ to point along the $-y$ direction.  

In the absence of pairing, TI surface states exhibit spin-momentum locking, i.e, a state on the Fermi surface defined by $|{\bm k}| = |\mu|/v_F \equiv k_F$ is spin-polarized along the in-plane direction perpendicular to its momentum. As a result, an in-plane magnetic field $\bm B$ displaces the entire Fermi surface in the perpendicular direction in the Brillouin zone. This is completely different from the case of an ordinary metal without SOC, where Zeeman field splits a spin-degenerate Fermi surface into two.

Now consider proximity-induced pairing in the presence of Zeeman field.
Importantly, when $|\bm B|$ is smaller than the upper critical field $B_{c2}$ that destroys the parent $s$-wave superconductor, the proximity-induced pairing potential $\Delta$ on the TI surface states remains finite.   Furthermore, we consider low temperature cases where the parent superconducting order parameter and hence induced pairing potential $\Delta$ do not have significant changes when external field $|\bm B|<B_{c2}$ is applied \cite{Douglass,Gupta,Nambu}. 
For weak field, the surface states remain fully gapped.  The proximity-induced gap is now anisotropic: the gap minimum is located at $\pm k_F \hat{\bm x}$, where the Zeeman field ${\bm B} \parallel \hat{\bm y}$ creates the largest energy difference between electrons of opposite momenta on the Fermi surface and thus has the strongest pair breaking effect. At $ V=\Delta$, the gap at $\pm k_F \hat{\bm x}$ closes.
For $V>\Delta$, the spectrum of $H$ becomes gapless and exhibits two Fermi surfaces of zero-energy Bogoliubov quasiparticles. The resulting phase diagram in terms of $|\bm B|$ is shown in Fig. \ref{1}a.

The quasiparticle spectrum of the Hamiltonian $H$ can be calculated analytically under the physically realistic condition $\Delta, V \ll |\mu|$. In this limit, we first diagonalize the TI surface Hamiltonian with $\Delta=V=0$, which we denote by $H_0$:
\begin{eqnarray}
H_0 =  \sum_{\bm k}   (v_F |\bm k| -\mu) f^\dagger_{{\bm k}} f_{{\bm k}}  + (-v_F |\bm k| -\mu) d^\dagger_{{\bm k}} d_{{\bm k}}
\end{eqnarray}
where $f^\dagger$ and  $d^\dagger$ are associated with conduction and valence bands respectively, defined by
\begin{eqnarray}
f^\dagger_{\bm k} = (c^\dagger_{\bm k \uparrow} + i e^{i \theta_{\bm k} } c^\dagger_{\bm k \downarrow} )/\sqrt{2}, \; \;
d^\dagger_{\bm k} = (c^\dagger_{\bm k \uparrow} - i e^{i \theta_{\bm k} } c^\dagger_{\bm k \downarrow})/\sqrt{2},  \nonumber
\end{eqnarray}
with $e^{i \theta_{\bm k}}\equiv (k_x + i k_y)/|\bm k|$.
We further rewrite the pairing and Zeeman term in the band basis using $f^\dagger_{\bm k}, d^\dagger_{\bm k}$. The Zeeman term now reads
\begin{eqnarray*}
H_{Z}&\equiv &c_{\bm k}^{\dagger}(-V\sigma_y)c_{\bm k}=iV(c_{\bm k\uparrow}^{\dagger}c_{\bm k\downarrow}-c_{\bm k\downarrow}^{\dagger}c_{\bm k\uparrow})\\
&=&-\frac{V}{|\bm k|}[k_{x}(f^\dagger_{\bm k}f_{\bm k}-d^\dagger_{\bm k}d_{\bm k})+ik_{y}(d^\dagger_{\bm k}f_{\bm k}-f^\dagger_{\bm k}d_{\bm k})]
\end{eqnarray*}
which involves momentum $ \bm k $ due to the spin-momentum locking. The pairing term now becomes
\begin{eqnarray*}
H_{P}&\equiv &\Delta( c^{\dagger}_{\bm k\uparrow}c^{\dagger}_{-\bm k\downarrow}-c^{\dagger}_{\bm k\downarrow}c^{\dagger}_{-\bm k\uparrow} +h.c.)\\
&=&\Delta [ie^{-i\theta_{\bm k}}(f^\dagger_{\bm k}f_{-\bm k}^{\dagger}-d^\dagger_{\bm k}d_{-\bm k}^{\dagger})+h.c.]
\end{eqnarray*}

Since only states in the vicinity of Fermi energy are strongly affected by pairing and Zeeman fields in the limit $\Delta, V \ll |\mu|$, for the purpose of solving the quasiparticle spectrum it suffices to keep terms involving conduction (valence) band operators only for $\mu>0$ ($\mu<0$). Assuming $\mu>0$, the original Dirac Hamiltonian $H$ after projection becomes a single-band model:
\begin{eqnarray}
H &\approx& \sum_{\bm k}  \epsilon_{\bm k} f^\dagger_{{\bm k}} f_{{\bm k}} - (V k_x/k_F)
 f^{\dagger}_{\bm k} f_{\bm k}    \nonumber \\
 &+& \frac{1}{2}\Delta (i e^{-i\theta_{\bm k}} f^\dagger_{\bm k} f^\dagger_{- \bm k} + h.c.)
 \label{oneband}
\end{eqnarray}
with $\epsilon_{\bm k} = v_F |\bm k| -\mu$.
Note that  the original $s$-wave pairing takes the form of a $(p_x + i p_y)$-like pairing in the reduced Hamiltonian \cite{FuKane}, while the Zeeman field takes the form of a vector potential $A_x = V/k_F$. We note that the equivalence bewteen an in-plane Zeeman field and a vector potential is exact for 2D Dirac Hamiltonian.

Diagonalizing Eq.(\ref{oneband}) yields the quasiparticle spectrum of $H$ near Fermi energy
\begin{eqnarray}
H &\approx &\sum_{\bm k} E_{\bm k} \gamma^\dagger_{\bm k} \gamma_{\bm k},\quad
\gamma_{\bm k}=u_{\bm k}f_{\bm k}^{\dagger}+u_{-\bm k}^{*}f_{-\bm k}\nonumber \\
E_{\bm k} &=& \sqrt{\epsilon_{\bm k}^2+\Delta^2 }- Vk_x/k_F \;
\textrm{ for } k \sim k_F,\label{E}
\end{eqnarray}
where the complex wavefunction reads
\begin{eqnarray}
u_{\bm k}=e^{\frac{i}{2}(\pi /2-\theta_{\bm k})}\frac{k_x}{|k_x|}\sqrt{\frac{1}{2}\left(1+\frac{k_{x}}{|k_x|}\frac{\epsilon_{\bm k}}{\sqrt{\epsilon_{\bm k}^2+\Delta^2}}\right)}.
\end{eqnarray}

The quasiparticle energy $ E_{\bm k} $ and its particle-hole partner $ -E_{-\bm k} $ at different Zeeman fields $ V $ are shown in Fig. \ref{1}a for the special case where $ k_y =0 $. It can be found that Zeeman field $V$ tilts the quasiparticle spectrum $ E_{\bm k} $: On one side $ (k_x >0) $ $ V $ lowers $ E_{\bm k} $ while on the other side $ (k_x<0) $ $ V $ increases $ E_{\bm k} $. This phenomenon will affect density of states of the system as will be discussed in the next section.

The Bogoliubov Fermi surface is thus given by $ E_{\bm k}=0 $, and one banana-shaped Fermi surface and its particle-hole partner are found located at two sides of the $k_y$-axis respectively as shown in Fig. \ref{1}b. Importantly, the partial Fermi surface $(k_x>0)$ is formed by electron and hole Fermi arcs. In terms of polar coordinate, the electron (+) and hole (-) arcs can be expressed as $|\bm k|= k_{F}\pm\sqrt{V^2\cos^2\theta -\Delta^2}/v_F $ where $\theta =\tan^{-1}(k_{y}/k_{x})$ is the polar angle of $\bm k$. It can be seen that along the original Fermi surface, two parts $ |\theta\pm\pi/2|<\pi/2-\theta_{m} $ are fully gapped while the rest parts $ |\theta|<\theta_{m} $ and $ |\pi -\theta|<\theta_{m} $ are paired together to form a new Bogoliubov Fermi surface, i.e. the partial Fermi surface. Here $ \theta_{m}=\cos^{-1}(\Delta/V) $.

The partial Fermi surface found here is robust and owes its existence to the sign change of the quasiparticle spectrum $E_{\bm k}$ in momentum space. A different type of Bogoliubov Fermi surface is theoretically shown to exist in certain centrosymmetric superconductors with unconventional pairings that breaks time-reversal symmetry \cite{Agterberg,Volovik}. In that case, the combination of particle-hole and inversion symmetry ensures that in Nambu space the quasiparticle spectrum at every $\bm k$ comes in pairs, $( E_{\bm k}, -E_{\bm k} )$. For certain pairings, the quasiparticle band crossing defined by $E_{\bm k} = -E_{\bm k} = 0$ leads to a Bogoliubov Fermi surface. However, this type of Bogoliubov Fermi surface is protected by inversion symmetry, and the band crossing generally becomes anti-crossing if inversion symmetry is broken. On the contrary, the systems considered here are non-centrosymmetric, and the partial Fermi surface we found is completely robust against all perturbations.

The charge and spin distributions of the partial Fermi surfaces are encoded in the wavefunction $u_{\bm k}$. We define the charge $ Q_{\bm k}\equiv -e\langle c^{\dagger}_{\bm k}c_{\bm k}-h^{\dagger}_{\bm k}h_{\bm k}\rangle $ and spin $\bm S_{\bm k}\equiv\langle c^{\dagger}_{\bm k}\bm\sigma c_{\bm k}-h_{\bm k}^{\dagger}\bm\sigma^{*}h_{\bm k}\rangle$ of the state $ \vert\gamma_{\bm k}\rangle\equiv\gamma_{\bm k}^{\dagger}\vert\text{GS}\rangle $, where $ h_{\bm k,s}=c_{-\bm k,s}^{\dagger} $ is the hole creation operator with momentum $\bm k$ and spin $s=\uparrow,\downarrow$, $ \vert\text{GS}\rangle $ denotes the ground state, and $ \langle\dots\rangle =\langle\gamma_{\bm k}\vert\dots\vert\gamma_{\bm k}\rangle$. It can be computed that
\begin{eqnarray}\label{Q}
\frac{Q_{\bm k}}{e}=-\frac{k_{x}}{|k_x|}\frac{\epsilon_{\bm k}}{\sqrt{\epsilon_{\bm k}^2+\Delta^2}},\quad
\bm S_{\bm k}=\frac{k_x\hat{\bm y}-k_y\hat{\bm x}}{|\bm k|}.
\end{eqnarray}

The charge and spin distributions of partial Fermi surface are shown in Fig. \ref{1}b. It is found that along the same $ \bm k $ direction, quasiparticles at the electron and hole arcs have opposite charges but the same spin. In fact in terms of polar angle $\theta$, the charge $ Q $ and spin $ \bm S $ distributions of electron (+) and hole $(-)$ arcs are $ Q_{\pm}(\theta)=\mp e\sqrt{\cos^2\theta -\cos^2\theta_{m}} $ and $ \bm S_{\pm}(\theta)=(-\sin\theta,\cos\theta ,0) $. Hence the total charge integrated over the partial Fermi surface is zero. On the other hand, the spin-momentum locking of partial Fermi surface is the same as the original electron Fermi surface.

{\bf Density of States}---
The partial Fermi surface in the gapless superconducting phase leads to nontrivial features in the density of states. Recall that 
in a conventional 2D $s$-wave superconductor without SOC, the in-plane magnetic field below upper critical field will uniformly split Bogoliubov quasiparticle spectrum into two by the amount of Zeeman energy. Thus the density of states (DOS) in this case is the superposition of two shifted BCS-type DOS \cite{Moodera}.

For TI surface states with spin-nondegenerate Fermi surfaces and strong spin-momentum locking in its normal phase, instead of splitting Fermi surfaces, Zeeman field behaves as the vector potential. In the superconducting phase, Zeeman field changes the gap size and eventually creates gapless Bogoliubov quasiparticles as shown in Fig. \ref{1}a. As a result, the DOS will be qualitatively different from that of conventional superconductors under Zeeman field.

\begin{figure}
\centering
\includegraphics[width=3.2in]{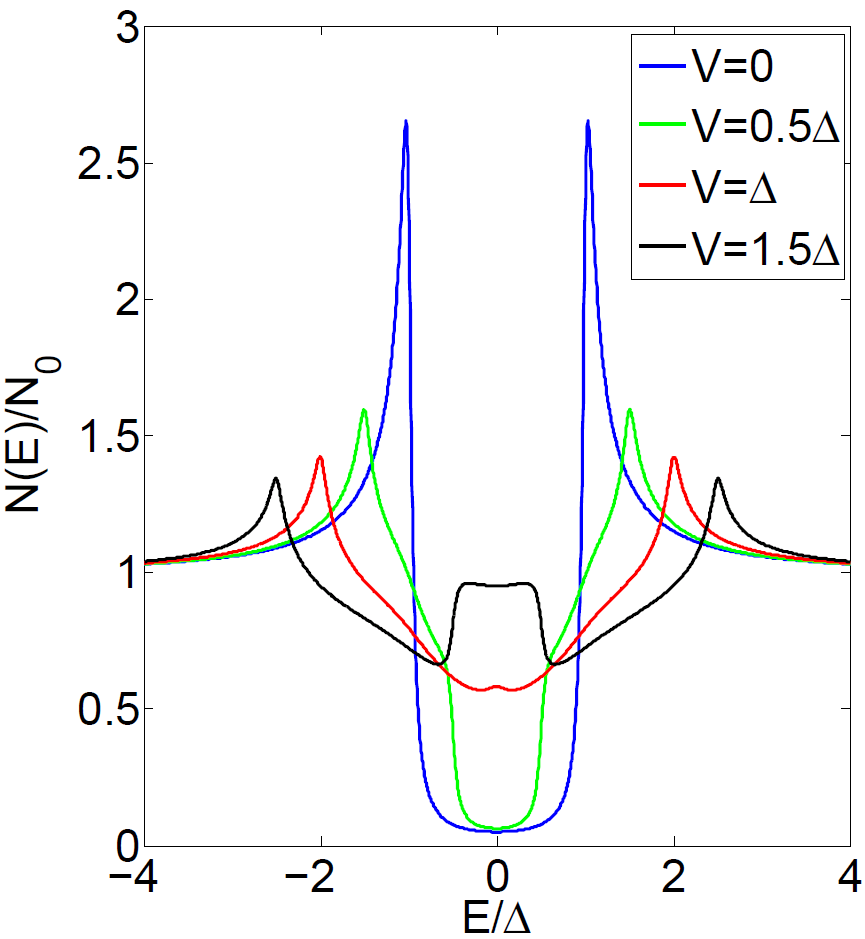}
\caption{Density of states (DOS) $ N(E) $ as functions of energy $E$ with Zeeman energy $V=0,0.5\Delta ,\Delta$ and $1.5\Delta$, and chemical potential $ \mu =90\Delta $. Here $ N_0 $ is the normal state DOS and $\Delta$ is the pairing amplitude. Other parameters are the same as Fig. \ref{1}b.}\label{2}
\end{figure}

In Fig. \ref{2}, we numerically calculate the DOS $ N(E)=-\text{Im}[\text{tr}\mathcal{G}(E,\bm k)]/\pi $ of the gapless superconducting phase, normalized by the normal state DOS $ N_{0}=2\pi\mu/v_{F}^2 $, where $ \mathcal{G}(E,\bm k) $ is Gor'kov Green's function. It can be found that when no field is applied $V=0$, the conventional BCS-type DOS is found for $N(E)$ with energy gap and coherence peak at the same position $ E=\Delta $. When $ V $ increases the energy gap decreases as $ \Delta -V $ until $ V\geqslant\Delta $ while the energy of coherence peak increases as $ \Delta +V $. In quasiparticle spectra of Fig. \ref{1}a, the energy gap corresponds to energy $ E(k_{F}\hat{\bm x})=\Delta -V $ while the coherence peak corresponds to energy $ E(-k_F\hat{\bm x})=\Delta +V $.

When $V=\Delta$, the system becomes nodal at $\pm k_F\hat{\bm x}$. Near the nodal point $ k_F\hat{\bm x} $ we have $ E(\bm p+k_F\hat{\bm x})=\frac{1}{2}v_{F}^2(p_{x}^2/\Delta +\Delta p_{y}^2/\mu^2) $, and hence close to zero energy the DOS $ N(E)=N_{0}/2 $ is a constant as shown in Fig. \ref{2}.

For larger magnetic field $ V>\Delta $, the system is gapless with partial Fermi surface. The partial Fermi surface has a much smaller $\bm k$-space area than the original Fermi surface, thus far from zero energy partial Fermi surface can be regarded as point nodes, and $N(E)$ behaves linearly in $E$ when $V-\Delta <E<V+\Delta$, as shown in Fig. \ref{2}. Close to zero energy, $ N(E) $ shows a plateau with height $N_0$ and width $V-\Delta$, due to states near partial Fermi surface.

Fig. \ref{2} is obtained by assuming the proximity-induced pairing $ \Delta $ is not affected by Zeeman energy $ V $. This assumption is justified at low temperatures for magnetic fields smaller than the upper critical field \cite{Douglass,Gupta,Nambu}. For higher temperatures, the induced pairing order parameter $ \Delta $ and hence DOS may change according to details of the whole system.

Our results of TI surface states turn out to be quite general and can be applied in 2D superconductors with Rashba and even general SOC.

{\bf Two-Dimensional Rashba Systems}---
Recently, 2DEG with Rashba SOC and induced pairing has been realized experimentally in quantum well systems such as the Al/InAs heterostructures \cite{Marcus0,Marcus1,Marcus2}. The induced superconducting phase of such systems in the presence of in-plane magnetic fields can also host gapless Bogoliubov quasiparticles.

In the 2DEG with Rashba SOC there are two spin-split Fermi surfaces in the normal phase, of which each one is spin-momentum locked, similar to that of TI surface states. When Rashba splitting energy is much larger than the Zeeman energy and induced pairing, the couplings between two Fermi surfaces can be neglected and we can regard the system as two copies of TI surface states. As a result, two partial Fermi surfaces in the gapless superconducting phase will be reconstructed from two original Fermi surfaces respectively. And the total DOS will be the sum of DOS from individual Fermi surfaces. When the $g$-factors and induced pairing potentials are the same for two Fermi surfaces, the normalized total DOS will be the same as that of single Fermi surface, and the result in Fig. \ref{2} still applies.

For the typical Al/InAs heterostructure, the Rashba SOC energy is about 0.2 eV, $ g $-factor is $|g|\sim 10$, and the induced pairing is $ \Delta\sim $0.1 meV \cite{Marcus0}. Thus when in-plane field $|\bm B|\gtrsim$0.5 T the Zeeman energy $V$ can surpass $ \Delta $, and the gapless superconducting phase is realized with two partial Fermi surfaces.

{\bf General Spin-Orbit Coupling}---
The results for TI surface states and 2D Rashba systems can be genralized to 2DEG with induced pairing $ \Delta $, strong SOC with arbitrary form and general in-plane Zeeman field. To be specific, consider the following  Hamiltonian
\begin{eqnarray}\label{G}
H&=&\sum_{\bm k}c^{\dagger}_{\bm k}[\xi_{\bm k}\sigma_0 +\bm g_{\bm k}\cdot\bm\sigma +\bm V\cdot\bm\sigma]c_{\bm k} \nonumber \\
&+& \Delta( c^{\dagger}_{\bm k\uparrow}c^{\dagger}_{-\bm k\downarrow} +h.c.),
\end{eqnarray}
where $ \xi_{\bm k}=|\bm k|^2/2m-\mu $ is the kinetic energy with effective mass $m$ and chemical potential $\mu$, $\bm g_{\bm k}=-\bm g_{-\bm k}$ is the SOC vector and $ \bm V =g\mu_B \bm B $ is the Zeeman field induced by in-plane field $\bm B$. If $ \bm g_{\bm k}\propto (-k_y,k_x,0) $ the Hamiltonian describes the 2D Rashba system, and if in addition $ m\to\infty ,\bm V =-V\hat{\bm y}$ the Hamiltonian becomes (\ref{H}) for TI surface states.

In general the Hamiltonian (\ref{G}) yields two bands $ \epsilon_{\bm k,\pm}=\xi_{\bm k}\pm|\bm g_{\bm k}| $ and hence inner and outer Fermi surfaces in the normal phase $ V=\Delta =0 $. When the SOC energy splitting $ \epsilon_{\text{SO}}=\text{min}_{\xi_{\bm k}=0}|\bm g_{\bm k}| $ between two Fermi surfaces is much larger than $\Delta$, we can treat the two Fermi surfaces separately.

Without loss of generality we focus on the states near inner Fermi surface $ \epsilon_{\bm k,+}=0 $ and apply Zeeman field $\bm V$ and induce pairing $\Delta$. To start with, we consider the $\bm k$ points where $ \bm g_{\bm k}\parallel\bm V $ and $ \bm g_{\bm k}\perp\bm V $.

For $\bm k$ points where $ \bm g_{\bm k} $ is anti-parallel to $\bm V$, under Zeeman field $\bm V$ the electron state with energy $ \epsilon_{\bm k,+} $ will be shifted to $ \epsilon_{\bm k,+}-|\bm V| $, and its time-reversal hole state at $-\bm k$ with energy $ -\epsilon_{\bm k,+} $ will be shifted to $ -\epsilon_{\bm k,+}-|\bm V| $. When pairing $\Delta$ is induced, the electron and hole states will form a Bogoliubov quasiparticle with energy $ \sqrt{\epsilon_{\bm k,+}^2+\Delta^2}-|\bm V| $, shifted by Zeeman energy $ |\bm V| $, just as conventional superconductors without SOC. When $ |\bm V|>\Delta $ these quasiparticles will become gapless.

Unlike the previous case, when $ \bm g_{\bm k}\perp\bm V $, the electron state at $ \bm k $ with energy $ \epsilon_{\bm k,+} $ will be changed to $ \mathcal{E}_{\bm k,+}=\xi_{\bm k}+\sqrt{|\bm g_{\bm k}|^2+|\bm V|^2} $, and its time-reversal hole state at $-\bm k$ with energy $ -\epsilon_{\bm k,+} $ will be changed to $ -\mathcal{E}_{\bm k,+}$. Thus in the superconducting phase the Bogoliubov quasiparticles formed by these states always have the gapped spectrum $ \sqrt{\mathcal{E}_{\bm k,+}^2+\Delta^2} $ up to leading order in Zeeman field \cite{Yanase1,Yanase2}.

For the general $ \bm k $ points, under physical conditions $ \Delta,|\bm V|\ll\epsilon_{\text{SO}} $ we can work out the quasiparticle spectrum up to the first order in $|\bm V|$:
\begin{eqnarray}
E_{\bm k}=\sqrt{\epsilon_{\bm k,+}^2+\Delta^2}+\bm V\cdot\bm g_{\bm k}/|\bm g_{\bm k}|,
\end{eqnarray}
which is generalization of (\ref{E}). Thus from this spectrum, the partial Fermi surface formed by electron and hole Fermi arcs can be worked out. As TI surface states and 2D Rashba systems, the electron and hole arcs have opposite charge distributions and the same spin texture.

{\bf Conclusion}---
In this work, we show that an in-plane magnetic field can drive the superconducting 2DEG with strong in-plane SOC such as TI surface states into the gapless superconducting phase, where a special type of Bogoliubov Fermi surface called partial Fermi surface is found. Reconstructed from ungapped part of the original electron Fermi surface, the partial Fermi surface is formed by electron and hole Fermi arcs whose charge distributions are opposite while spin textures are the same. In terms of DOS, we predict that with increasing Zeeman field, the energy gap will decrease to zero while the energy of coherence peak will increase, which are both linear in Zeeman energy. Properties of partial Fermi surface can be further probed by quasiparticle interference measurements under an in-plane magnetic field, and can reveal useful information about the spin textures of electron Fermi surface in the normal state.

{\bf Acknowledgement}---
We thank Ali Yazdani for interesting discussions. This work is supported by DOE Office of Basic Energy Sciences, Division of Materials Sciences and Engineering under Award DE-SC0010526. NFQY acknowledges the support of HKRGC through C6026-16W.


\begin{thebibliography}{99}
\bibitem{FuKane} L. Fu and C. L. Kane, Phys. Rev. Lett. \textbf{100}, 096407 (2008).
\bibitem{Alicea} J. Alicea, Phys. Rev. B \textbf{81}, 125318 (2010).
\bibitem{Oreg} Y. Oreg, G. Refael, and F. von Oppen, Phys. Rev. Lett. \textbf{105}, 177002 (2010).
\bibitem{Lutchyn} R. M. Lutchyn, J. D. Sau, S. Das Sarma, Phys. Rev. Lett. \textbf{105}, 077001 (2010).
\bibitem{Potter} A. C. Potter and P. A. Lee, Phys. Rev. B \textbf{83}, 094525 (2011).
\bibitem{Kouwenhoven}  V. Mourik, K. Zuo, S. M. Frolov, S. R. Plissard, E. P. A. M. Bakkers, and L. P. Kouwenhoven, Science \textbf{336}, 1003 (2012).
\bibitem{ADas} A. Das, Y. Ronen, Y. Most, Y. Oreg, M. Heiblum, H. Shtrikman, Nat. Phys. \textbf{8}, 887 (2012).
\bibitem{JFJia} H.-H. Sun, K.-W. Zhang, L.-H. Hu, C. Li, G.-Y. Wang, H.-Y. Ma, Z.-A. Xu, C.-L. Gao, D.-D. Guan, Y.-Y. Li, C. Liu, D. Qian, Y. Zhou, L. Fu, S.-C. Li, F.-C. Zhang, and J.-F. Jia, Phys. Rev. Lett. \textbf{116}, 257003 (2016).
\bibitem{JMLu} J. M. Lu, O. Zeliuk, I. Leermakers, N. F. Q. Yuan, U. Zeitler, K. T. Law, J. T. Ye, Science \textbf{350}, 1353 (2015).
\bibitem{Saito} Y. Saito, Y. Nakamura, M. S. Bahramy, Y. Kohama, J. Ye, Y. Kasahara, Y. Nakagawa, M. Onga, M. Tokunaga, T. Nojima, Y. Yanase, Y. Iwasa, Nat. Phys. \textbf{12}, 144–149 (2016).
\bibitem{Xi} X. Xi, Z. Wang, W. Zhao, J.-H. Park, Kam Tuen Law, Helmuth Berger, László Forró, Jie Shan, Kin Fai Mak, Nat. Phys. \textbf{12}, 139–143 (2016).
\bibitem{BTZ} B. T. Zhou, N. F. Q. Yuan, H.-L. Jiang, and K. T. Law, Phys. Rev. B \textbf{93}, 180501(R) (2016).
\bibitem{FuBerg} L. Fu and E. Berg, Phys. Rev. Lett. \textbf{105}, 097001 (2010).
\bibitem{Sho} S. Nakosai, Y. Tanaka, and N. Nagaosa, Phys. Rev. Lett. \textbf{108}, 147003 (2012).
\bibitem{KoziiFu} V. Kozii and L. Fu, Phys. Rev. Lett. \textbf{115}, 207002 (2015).
\bibitem{YXWang} Y. Wang, G. Y. Cho, T. L. Hughes, and E. Fradkin, Phys. Rev. B \textbf{93}, 134512 (2016).
\bibitem{WuMartin} F. Wu and I. Martin, Phys. Rev. B {\bf 96}, 144504 (2017).
\bibitem{Clogston} A. M. Clogston, Phys. Rev. Lett. \textbf{9}, 266 (1962).
\bibitem{Chandrasekhar} B. S. Chandrasekhar, Appl. Phys. Lett. \textbf{1}, 7 (1962).
\bibitem{Maki} K. Maki and T. Tsuneto, Prog. Theor. Phys. \textbf{31}, 6 (1964).
\bibitem{Gorkov} V. Barzykin and L. P. Gor’kov, Phys. Rev. Lett. \textbf{89}, 227002 (2002).
\bibitem{Olga} O. Dimitrova and M. V. Feigel’man, Phys. Rev. B \textbf{76}, 014522 (2007).
\bibitem{Patrick} K. Michaeli, A. C. Potter, and P. A. Lee, Phys. Rev. Lett. \textbf{108}, 117003 (2012).
\bibitem{MXWang} M.-X. Wang, C. Liu, J.-P. Xu, F. Yang, L. Miao, M.-Y. Yao, C. L. Gao, C. Shen, X. Ma, X. Chen, Z.-A. Xu, Y. Liu, S.-C. Zhang, D. Qian, J.-F. Jia, Q.-K. Xue, Science  \textbf{336}, 52 (2012).
\bibitem{SYXu} S.-Y. Xu, N. Alidoust, I. Belopolski, A. Richardella, C. Liu, M. Neupane, G. Bian, S.-H. Huang, R. Sankar, C. Fang, B. Dellabetta, W. Dai, Q. Li, M. J. Gilbert, F. Chou, N. Samarth, M. Zahid Hasan, Nat. Phys. \textbf{10}, 943 (2014).
\bibitem{Hart} S. Hart, H. Ren, T. Wagner, P. Leubner, M. Mühlbauer, C. Brüne, H. Buhmann, L. W. Molenkamp, A. Yacoby, Nat. Phys. \textbf{10}, 638 (2014).
\bibitem{Yazdani} Private communications with A. Yazdani.
\bibitem{Marcus0} J. Shabani, M. Kjaergaard, H. J. Suominen, Y. Kim, F. Nichele, K. Pakrouski, T. Stankevic, R. M. Lutchyn, P. Krogstrup, R. Feidenhans'l, S. Kraemer, C. Nayak, M. Troyer, C. M. Marcus, and C. J. Palmstrom, Phys. Rev. B \textbf{93}, 155402 (2016).
\bibitem{Marcus1} H. J. Suominen, J. Danon, M. Kjaergaard, K. Flensberg, J. Shabani, C. J. Palmstrom, F. Nichele, C. M. Marcus, Phys. Rev. B \textbf{95}, 035307 (2017).
\bibitem{Marcus2} H. J. Suominen, M. Kjaergaard, A. R. Hamilton, J. Shabani, C. J. Palmstrom, C. M. Marcus, F. Nichele, Phys. Rev. Lett. \textbf{119}, 176805 (2017).
\bibitem{Douglass} D. H. Douglass, Jr., Phys. Rev. Lett. \textbf{6}, 346 (1961).
\bibitem{Gupta} K. K. Gupta and V. S. Mathur, Phys. Rev. \textbf{121}, 107 (1961).
\bibitem{Nambu} Y. Nambu and S. F. Tuan, Phys. Rev. \textbf{128}, 2622 (1962).
\bibitem{Agterberg} D. F. Agterberg, P. M. R. Brydon, and C. Timm, Phys. Rev. Lett. \textbf{118}, 127001 (2017).
\bibitem{Volovik} G.E. Volovik,  Phys. Lett. A {\bf 142}, 282-284 (1989).
\bibitem{Moodera} E. Strambini, V. N. Golovach, G. De Simoni, J. S. Moodera, F. S. Bergeret, and F. Giazotto, Phys. Rev. Materials \textbf{1}, 054402 (2017).
\bibitem{Yanase1} A. Daido and Y. Yanase, Phys. Rev. B \textbf{94}, 054519 (2016). 
\bibitem{Yanase2} A. Daido and Y. Yanase, Phys. Rev. B \textbf{95}, 134507 (2017).
\end{thebibliography}
\end{document}